\documentclass[sigconf]{acmart}

\definecolor{revised}{rgb}{0,0,0}
\definecolor{ty}{rgb}{0,0,0}

\AtBeginDocument{%
  \providecommand\BibTeX{{%
    \normalfont B\kern-0.5em{\scshape i\kern-0.25em b}\kern-0.8em\TeX}}}

\copyrightyear{2022}
\acmYear{2022}
\setcopyright{acmcopyright}\acmConference[MET'22]{7th International Workshop on
Metamorphic Testing}{May 9, 2022}{Pittsburgh, PA, USA}
\acmBooktitle{7th International Workshop on Metamorphic Testing (MET'22), May 9,
2022, Pittsburgh, PA, USA}
\acmPrice{15.00}
\acmDOI{10.1145/3524846.3527341}
\acmISBN{978-1-4503-9307-2/22/05}

\begin{document}

\title{Testing Ocean Software with Metamorphic Testing}

\author{Quang-Hung Luu}
\affiliation{%
  \institution{Swinburne University of Technology}
  \city{Hawthorn}
  \state{Victoria}
  \country{Australia}
  \postcode{3122}
}
\email{hluu@swin.edu.au}

\author{Huai Liu}
\affiliation{%
  \institution{Swinburne University of Technology}
  \city{Hawthorn}
  \state{Victoria}
  \country{Australia}
  \postcode{3122}
}
\email{hliu@swin.edu.au}

\author{Tsong Yueh Chen}
\affiliation{%
  \institution{Swinburne University of Technology}
  \city{Hawthorn}
  \state{Victoria}
  \country{Australia}
  \postcode{3122}
}
\email{tychen@swin.edu.au}

\author{Hai L. Vu}
\affiliation{%
  \institution{Monash University}
  \city{Clayton}
  \state{Victoria}
  \country{Australia}
  \postcode{3800}
}
\email{hai.vu@monash.edu}

\renewcommand{\shortauthors}{Q.-H.~Luu et al.}

\begin{abstract}
Advancing ocean science has a significant impact to the development of the world, from operating a safe navigation for vessels to maintaining a healthy and diverse ocean ecosystem. Various ocean software systems have been extensively adopted for different purposes, for instance, predicting hourly sea level elevation across shorelines, simulating large-scale ocean circulations, as well as integrating into Earth system models for weather forecasts and climate projections. Regardless of their significance, guaranteeing the trustworthiness of ocean software and modelling systems is a long-standing challenge. The testing of ocean software suffers a lot from the so-called oracle problem, which refers to the absence of test oracles mainly due to the nonlinear interactions of multiple physical variables and the high complexity in computation. In the ocean, observed tidal signals are distorted by non-deterministic physical variables, hindering us from knowing the ``true'' astronomical tidal constituents existing in the timeseries. In this paper, we present how to test tidal analysis and prediction (TAP) software based on metamorphic testing (MT), a simple yet effective testing approach to the oracle problem. In particular, we construct metamorphic relations from the periodic property of astronomical tide, and then use them to successfully detect a real-life defect in an open-source TAP software. We also conduct a series of experiments to further demonstrate the applicability and effectiveness of MT in the testing of TAP software. Our study not only justifies the potential of MT in testing more complex ocean software and modelling systems, but also can be expanded to assess and improve the quality of a broader range of scientific simulation software systems.
\end{abstract} 




\maketitle

\section{Introduction}

Our Earth is actually a world of water as over 70\% of its surface is covered by the ocean. Ocean science, also being referred to as marine science or oceanography, is an important discipline of the Earth sciences, covers a wide range of topics, from ocean circulation, waves, storm surge, sea level rise, heat content, water transport to ecosystem dynamics, ocean acidification and plate tectonics of the sea bed \cite{garrison2016}. In the current computing age, various software systems have been developed to support the ocean modelling and computations, which help our scientists simulate, model and predict ocean conditions, and thus assist the relevant industries and human activities as well as contribute to advance other disciplines \cite{roisin2011}.

Being a dynamic of the ocean, astronomical tide is the periodic variability of sea levels on the Earth exerted by gravitational forces of the Moon and the Sun. We are hence unable to observe the pure astronomical tide in sea level elevations. Instead, sea level elevations are a result of co-existence and interaction between tides and other non-deterministic variables such as surface wind, atmospheric pressure and ocean currents at multiple spatio-temporal scales \cite{luu2015}. In other words, it is a great challenge to correctly obtain the ``pure'' astronomical tidal constituents  from sea level records. 

Tidal analysis and prediction (TAP) software is an ocean software and modelling system. It is developed to decompose tidal constituents from timeseries (i.e., series of data points arranged in the time order) of the sea level elevation, and adopt them to predict future sea level \cite{foreman1989,pawlowicz2002,codiga2011}. Having correct astronomical tidal constituents is a critical task because it contributes to assure a safe vessel navigation, validate complex ocean modelling systems \cite{luu2011,paduan2013}, estimate the rate of sea level rise \cite{tkalich2013,wahl2015,luu2015}, and even support the claims on oceanographic territories \cite{lyons2018}. On one hand, it is difficult to isolate pure astronomical tidal constituents. On the other hand, the TAP software is prone to defects and bugs. Because of both physical and computational reasons, it is difficult to determine whether the outcome from the TAP software is correct. 

It is a prevalent problem in many ocean software and modelling systems, including TAP software, that there does not exist a systematic mechanism, namely the \textsl{oracle}, to check the correctness of the computation result given any program input. Such a problem is termed as the \textsl{oracle problem} in the context of software testing. Among all testing techniques, \textsl{metamorphic testing} (MT) \cite{chen1998first,segura2016,chen2017} has been demonstrated as very effective in addressing the oracle problem which has helped reveal many longstanding real-life bugs in a wide variety of software systems. The core component of MT is a set of \textsl{metamorphic relations} (MRs), which are basically the necessary properties of the target algorithm, represented as the relationships among multiple inputs and their corresponding expected outputs. Numerous studies \cite{segura2016,chen2017,segura2018b} have shown that if well-defined, even a simple MR could help detect non-trivial bugs in a complicated system that suffers from the oracle problem.

Several studies have been conducted to use MT to test the software for scientific computations and simulations. For instances, in testing simulations, MT has been successfully adopted to test agent-based (ABM) and discrete-event (DES) simulations \cite{olsen2019}, hybrid ABM and DES systems \cite{farhan2021}, health care simulation \cite{murphy2011}, web-enabled simulation \cite{ahlgren2021}, and simulator platform for self-driving cars \cite{zhou2019,seymour2021}. In testing scientific software, MT is an effective technique to detect faults in simulation programs for designing nuclear power plants \cite{he2020}, bioinformatics programs \cite{chen2009bio}, epidemiological models \cite{ramanathan2012,pullum2012}, chemical reaction networks for prototyping nano-scale molecular devices \cite{gerten2020}, matrix calculation programs \cite{rahman2018}, solvers for partial differential equations \cite{chen2002}, multiple linear regression software \cite{luu2021}, ocean modelling \cite{hiremath2021}, storm water management model systems \cite{lin2018}, machine learning-based hydro-logical models \cite{yang2021wrr},  Monte-Carlo computational programs \cite{ding2011,pesu2018}, serverless scientific applications \cite{lin2021}, as well as other types of scientific software \cite{lin2018,ding2019,peng2021,kanewala2019cse}. For example, He {\it et al.} \cite{he2020} has found 33 bugs in simulation programs that are used to design and analyze nuclear power plants in a study that adopts MT.

In this paper, we demonstrate how to test TAP software using MT. Seven MRs are identified and applied to test UTide, a real-life and open-source TAP software. The paper has the following three contributions:
\begin{itemize}
    \item Based on the fundamental features of TAP, we propose two strategies for identifying MRs that are particularly suitable for TAP software.
    \item The testing based on our MRs has revealed a real-life defect in UTide.
    \item A series of experimental studies have been conducted to demonstrate the applicability and effectiveness of MT in the testing of TAP software, and more broadly, ocean software.
\end{itemize}

The rest of the paper is organized as follows. In Section~\ref{sec:background}, we introduce the background information of the present work, including the basic concepts of MT and TAP. In Section~\ref{sec:method}, we present two strategies of identifying MRs for TAP software, based on which, we define seven MRs for this study. The settings for our experimental studies are presented in Section~\ref{sec:experiment}, while the experimental results are analyzed in Section~\ref{sec:result}. We further discuss the results in Section~\ref{sec:discuss} and the related work in Section~\ref{sec:related}. Finally, the paper is summarized in Section~\ref{sec:conclusion}.

\section{Background}
\label{sec:background}

\subsection{Metamorphic Testing}

Metamorphic testing (MT) was first proposed in 1990s as an approach for test case generation \cite{chen1998}. MRs are leveraged to transform some existing test cases into new test cases. The basic steps for MT are as follows:
\begin{enumerate}
    \item Identify necessary properties of the target program and represent them in the form of MRs, that is, the relations among multiple inputs and their corresponding expected outputs.
    \item Generate \textsl{source test cases} using other test case generation techniques.
    \item Execute the source test cases.
    \item Construct \textsl{follow-up test cases} based on MRs and the source test cases (and even their test results obtained in Step (3)).
    \item Execute the follow-up test cases.
    \item Check the test results in Steps (3) \& (5) against the corresponding MRs.
\end{enumerate}

Note that MT makes use of MRs, instead of a test oracle, to verify the test results, as shown in the above Step (6). If an MR is violated in Step (6), we can say that a defect has been revealed. In other words, MRs provide an alternative mechanism to the oracle for test result verification, thus addressing the oracle problem.

Let us look at the following example to further illustrate how MT works. Suppose that a program $P$ calculates the shortest path between two nodes $a$ and $b$ in an undirected graph $G$. One possible MR can be defined as follows: Given a node $c$ that appears in the shortest path between $a$ and $b$, the length of the shortest path between $a$ and $b$ should be equal to the sum of the length of the shortest paths between $a$ and $c$ and that between $c$ and $b$. Assume that we already have the source test case $(G, a, b)$. We execute this test case and select a node $c$ from the execution result $P(G, a, b)$ (that is, the shortest path composed of a series of nodes). Then, we can construct two follow-up test cases, $(G, a, c)$ and $(G, c, b)$, the execution results of which are $P(G, a, c)$ and $P(G, c, b)$, respectively. Finally, we check whether the relation $|P(G, a, b)| = |P(G, a, c)| + |P(G, c, b)|$ holds or not. If the relation does not hold, it means that the MR is violated, which, in turn, implies the detection of a defect. 

\subsection{Tidal Analysis and Prediction}

Consider a time series of sea level observations $y(t)$ where $t=t_1,t_2,\ldots,t_M$ and $M$ is an odd integer. Assume the sea level is a combination of $N$ astronomical tidal constituents with frequencies $\sigma_1,\sigma_2,\ldots,\sigma_N$, each of which is constructed from a unique combination of six Doodson numbers. These numbers are associated with the mean longitudes of the Moon, the Sun, the lunar perigee and the solar perigee, together with the negative of the longitude of the Moon's ascending node and the mean lunar time \cite{foreman1989}.
Tidal response model equation \cite{foreman1989,pawlowicz2002} is as follows:
\begin{equation}
\label{equation:responsemodel}
 \hat y(t) = \hat a_0 
    + \hat a_1 t
    +\sum_{k=1}^N \hat b_{k} e^{i\sigma_k t}
    +  \hat b_{-k} e^{-i\sigma_k t}
\end{equation}
where $\hat b_{-k}$ and $\hat b_{-k}$ are unknown (complex) amplitude and $i$ is the imaginary unit. Its conjunction form in real numbers is as follows:
\begin{equation}
\label{equation:responsemodel2}
 \hat y(t) = \hat a_0 
    + \hat a_1 t
    +\sum_{k=1}^N 
    \left(
    \hat B_{k} \cos{\sigma_k t}
    + \hat C_{k} \sin{\sigma_k t}
    \right)
\end{equation}
where $\hat B_{k}=(\hat b_{k}+\hat b_{-k})/2$ and $\hat C_{k}=i(\hat b_{k}-\hat b_{-k})/2$. Alternatively, the response model can also be represented by the following equation:
\begin{equation}
\label{equation:responsemodel3}
 \hat y(t) = \hat a_0 
    + \hat a_1 t
    +\sum_{k=1}^N 
    \hat A_{k} \cos{(\sigma_k t + \hat\phi_k)}
\end{equation}
where the amplitude $\hat A_{k}$ and the phase $\hat\phi_k$ of a tidal constituent $k$ are defined in terms of $\hat B_{k}$ and $\hat C_{k}$ as follows:
\begin{equation}
\begin{aligned}\label{equation:amp}
\hat A_{k} &= \sqrt{\hat B_{k}+\hat C_{k}}\\
\hat\phi_k &= \arctan{(-\hat C_{k}/\hat B_{k})}
\end{aligned}
\end{equation}

After the equation is solved, some corrections can be applied. First, the computed phases $\hat\phi_k (k=1,2,\ldots,N)$ are referenced to the ``Greenwich phase'', which is defined as the phase of the equilibrium response at $0^\circ$ longitude (the Greenwich meridian). Second, nodal or satellite corrections are applied if a latitude is specified in the configurable input; otherwise, it will use a default value. 

A process for TAP consists of two modes. In the analysis mode, the TAP software takes the input of raw timeseries of sea level elevation $y(t)$ at time steps $t=t_1,t_2,\ldots,t_M$. It will also use other inputs such as tidal frequencies, configuration for which algorithm to be used for fitting, the latitude of tide gauge or the adoption of a trend. If any of these inputs is not provided, it will be automatically determined by the TAP software, through either assigning a default value or computing it using relevant algorithms. For instance, frequencies of $N$ astronomical tidal constituents $\sigma_1,\sigma_2,\ldots,\sigma_N$ can be either given by users or determined by the TAP itself, at the discretion of the users. The output of analysis mode consists of tidal amplitudes $\hat A_1,\hat A_2,\ldots,\hat A_N$ and phases $\hat \phi_1,\hat \phi_2,\ldots,\hat \phi_N$ together with the intercept $\hat a_1$ and the trend $\hat a_2$. In the prediction mode, the TAP software uses the determined tidal constants $\hat A_1,\hat A_2,\ldots,\hat A_N$ and phases $\hat \phi_1,\hat \phi_2,\ldots,\hat \phi_N$ as well as $\hat a_1$ and $\hat a_2$ to predict the sea level elevation $\hat y(t)$ at a given time $t$ according to Equation~\ref{equation:responsemodel3}.

\section{Metamorphic Relations for TAP software}
\label{sec:method}

Considering the unique features of TAP software, we design two strategies to derive MRs. 
The first approach is to construct MRs based on the core mathematical framework for the TAP, namely regression. The second approach is to construct MRs based on the properties of tidal waves.

\subsection{Regression-based MRs}

We can reformulate Equation~\ref{equation:responsemodel2} in the matrix form as follows:
\begin{equation}
 \hat{\mathbf{y}}(t) = {\mathbf{X}}^T(t) \hat\beta^*
\end{equation}
where
\begin{equation}
\label{equation:denote}
\begin{aligned}
  {\mathbf{X}(t)} &=
\begin{bmatrix}
    1 & 1 & 1 & \ldots & 1 \\
    t_1 & t_2 & t_3 & \ldots  & t_M \\
    \cos{\sigma_1 t_1} & \cos{\sigma_1 t_2}  & \cos{\sigma_1 t_3} & \ldots & \cos{\sigma_1 t_M} \\
    \sin{\sigma_1 t_1} & \sin{\sigma_1 t_2}  & \sin{\sigma_1 t_3} & \ldots & \sin{\sigma_1 t_M} \\
    \ldots & \ldots & \ldots  & \ldots  &\ldots \\
    \cos{\sigma_N t_1} & \cos{\sigma_N t_2}  & \cos{\sigma_N t_3} & \ldots & \cos{\sigma_N t_M} \\
    \sin{\sigma_N t_1} & \sin{\sigma_N t_2}  & \sin{\sigma_N t_3} & \ldots & \sin{\sigma_N t_M}   
\end{bmatrix}
    ; \\
  \hat{\beta}^* &=
\begin{bmatrix}
    \hat{ a_0} &
    \hat{ a_1} &
    \hat{B}_1 &
    \hat{C}_1 &
    \ldots &
    \hat{B}_N &
    \hat{C}_N
\end{bmatrix}
  ; \\
  \hat{\mathbf{y}}(t) &=
\begin{bmatrix}
    \hat{y}({t_1}) &
    \hat{y}({t_2}) &
    \ldots &
    \hat{y}({t_M})
\end{bmatrix}
\end{aligned}
\end{equation}

This equation can be solved using the ordinary least square (OLS) regression, which gives us the unique solution
\begin{equation}
\label{equation:beta}
 \hat\beta^* = ({\mathbf{X}}{\mathbf{X}}^T)^{-1} {\mathbf{X}}{\mathbf{y}}(t)
\end{equation}
where
\begin{equation}
  \mathbf{y}(t) =
\begin{bmatrix}
    {y}({t_1}) &
    {y}({t_2}) &
    \ldots &
    {y}({t_M})
\end{bmatrix}
\end{equation}

In the analysis, the output coefficient $\hat\beta$ is commonly used as the output which is defined as follows
\begin{equation}
\label{equation:beta2}
 \hat\beta = \begin{bmatrix}
    \hat{ a_0} &
    \hat{ a_1} &
    \hat{A}_1 &
    \hat{\phi}_1 &
    \ldots &
    \hat{A}_N &
    \hat{\phi}_N
\end{bmatrix}
\end{equation}
where the coefficients of $\hat\beta$ are derived from the coefficients of $\hat\beta^*$ according to Equation~\ref{equation:amp} and Equation~\ref{equation:denote}. 

\vspace{0.2cm}
Hereby, we reuse some MRs that have been developed in our previous study for the multiple linear regression (MLR) \cite{luu2021}. 
Given the source test case $I_s = \{(t_1^s,y_1^s),(t_2^s,y_2^s),\ldots,(t_M^s,y_M^s)\}$ and the corresponding output $O_s = \{ \hat\beta_s \}$. It should be noted that regression and TAP software are designed to solve different problems though they are based on the same mathematical approach. Therefore, they have some common MRs but also some different MRs. For example, we have an MR related to swapping or rotating two timeseries (i.e., the data for two variables) in the regression, but such kind of MRs cannot be defined for the TAP software. In this study, five MRs are reused to the TAP software as described as follows.

{\bf MR1}: {\it Inserting a predicted point.} Assume $(t_p,y_p)$ is a predicted point using Equation~\ref{equation:responsemodel2}. The follow-up test case can be constructed by $I_f = I_s \bigcup (t_p,\hat y_p)$, and the follow-up output should be $O_f = \{ \hat\beta_f \}$. Thus, this MR should have the relation $\hat\beta_f=\hat\beta_s$.

{\bf MR2}: {\it Reflecting the sea level.} The follow-up test case can be constructed by having $I_f = -I_s$, then $\hat\alpha_0^f=-\hat\alpha_j^0$ and $\hat\alpha_1^f=-\hat\alpha_j^1$. When there is only one astronomical tidal constituent $(N=1)$, the relation should also consist of $\hat A_1^f= \hat A_1^s$.

{\bf MR3}: {\it Shifting the sea level.} The follow-up test case can be constructed by having $I_f = I_s+h$, where $h$ is a real number, and then the intercept is shifted accordingly whilst other components of $\hat\beta$ are unchanged; that is, $\hat\alpha_0^f=\hat\alpha_0^s+h$ and $\hat\alpha_1^f=\hat\alpha_1^s$, $\hat A_k^f= \hat A_k^s, \hat\phi_k^f = \hat\phi_k^s$ for $k=1,2,\ldots,N$.

{\bf MR4}: {\it Scaling the sea level.} The follow-up test case can be constructed by $I_f = \gamma I_s$, where $ \gamma$ is a real number, and then the amplitude-related components of the follow-up output is scaled proportionally, that is, $\hat\alpha_0^f=\gamma \hat\alpha_0^s$, $\hat\alpha_1^f=\gamma \hat\alpha_1^s$, $\hat A_k^f= \gamma\hat A_k^s$ for $k=1,2,\ldots,N$.

{\bf MR5}: {\it Swapping samples.} If we swap two points in the input, then the follow-up output is the same, that is, $\hat\beta_f=\hat\beta_s$.

\subsection{Domain-specific MRs}

{\bf MR6}: {\it Cancellation of signals.} 
Consider the sea level elevation induced by a single tidal constituent component without a trend, that is, we have
\begin{equation}\label{equation:mr6}
 \hat y(t) = \hat a_0 
    + \hat A_{1} \cos{(\sigma_1 t + \hat\phi_1)}
\end{equation}
In the follow-up test case, the signal is cancelled by the predicted component, $y^f = y^s - \hat A_{1}^s \cos{(\sigma_1 t + \hat\phi_1^s)}$. Then, we should have $\hat A_1^f=0$.
MR6 comes from the intuition that if we suppress the oscillation of a tidal constituent,  then this wave component is no longer part of the timeseries.

{\bf MR7}: {\it Periodic shift of signals.} Consider the sea level elevation induced by a single tidal constituent with a trend, that is
\begin{equation}\label{equation:mr7}
 \hat y(t) = \hat a_0 
    + \hat a_1 t
    + \hat A_{1} \cos{(\sigma_1 t + \hat\phi_1)}
\end{equation}
In the follow-up test case, the timeseries $t^f_1,t^f_2,\ldots,t^f_M$ are all shifted by a magnitude of $2\pi/\sigma_1$, that is, $t^f_j=t^s_j+2\pi/\sigma_1$ for $j=1,2,\ldots,M$. 
The follow-up output is then expected to be $\hat a_0^f=\hat a_0^s-2\pi \hat a_1^s/\sigma_1$, and $\hat a_1^f=\hat a_1^s, \hat A_1^f=\hat A_1^s, \hat \phi_1^f=\hat \phi_1^s$. 
The rationale for MR7 is the periodic property of a tidal wave which requires the tide to be the same after a full period.

\section{Experiments}
\label{sec:experiment}

\subsection{Research questions}

In this study, we attempt to answer the following three research questions (RQs):

{\bf RQ1}: {\it Can MT help reveal any real-life defect?} We will examine whether or not the proposed MRs are able to detect any issue in the system under test (SUT). 

{\bf RQ2}: {\it Is MT applicableto detect different types of faults?} We will carry out the mutation analysis to further study the effectiveness of MT.

{\bf RQ3}: {\it Is there any difference in the effectiveness of MRs?} Its answer may hint us for the selection of the best MRs for testing.  

RQ1 will be addressed by testing the original program; whilst RQ2 and RQ3 will be answered based on the mutation analysis.

\subsection{System under test and mutation analysis}

UTide (Unified Tidal Analysis and Prediction) software developed by Wesley Bowman (https://github.com/wesleybowman/UTide) was used as our system under test. Its Python version has 2940 lines of code (LOC), taking the raw input of timeseries of sea level, and producing the output in the form of either tidal harmonic constants (in the analysis mode) or timeseries of tidal elevation (in the prediction mode). 
Having the latest version (v0.2.6) published one year ago, the software is stable after 4 years of development, receiving GitHub stars from 75 users and being forked by 44 developers.
It was configured to run on iMac M1 computer with Python 3 under an Anaconda environment. 

Mutation analysis is a commonly used method to examine the effectiveness of testing approaches \cite{jia2011}. In the mutation analysis, the original program will be modified by seeding a fault into a selected statement. Each modified version is referred to as a mutant, which should be unique to avoid double counts. It contains the code and behaviour that are not the same as the original program. If the fault seeded in a mutant can be detected using a test suite, it is said that the mutant is ``killed'' by the test suite. The effectiveness of the test suite (and hence the technique that generates the test suite) is measured by the percentage of mutants that they kill out of all considered mutants. The advantage of mutation analysis is to generate a wide range of faults to challenge the effectiveness of the newly developed testing method. 
In this study, we adopted the mutant-generating tool developed by Luu {\it et al.} \cite{luu2021}, and modified it to suit our purpose of testing Python-based programs.

Our focus is to test the core component of UTide. Hence, for the mutation analysis, we generated mutants from the core analysis component of the TAP. It contains 206 LOC (in ``\_solve.py'' from line 232 to line 438). Among mutation operators defined in \cite{luu2021} for C/C++ programs, we used five sets to generate mutants for the Python program, as presented in Table~\ref{table:mutants}. We then applied a filtering process to eliminate the equivalent mutants (which refers to the mutants that have the same execution behaviors as the original program): If a mutant always delivers the same output as the UTide on a set of randomly generated test cases, that mutant would be removed. We also manually inspected the remaining mutants to ensure that the non-equivalent mutants are unique. 
Finally, we obtained 38 non-equivalent mutants that are non-trivial (i.e., not killed by the Python interpreter) to be used for testing (Table~\ref{table:mutants}).

\begin{table*}[tbp] 
\vspace{0.2cm}
\caption{Summary of generated mutants and mutants used for testing, being classified by mutation operators.}
\label{table:mutants}
\begin{center}
\begin{tabular}{|l|l|l|r|}
\hline
\textbf{Mutation category} 
	& \textbf{mutation operators}
	& \textbf{Description}
	& \textbf{Number of used mutants}\\
\hline
    array\_index &
       [], [1+], [1-]   & Misplaced array index &  7      \\ 
    condition\_if &
        if(), if None, if False, if True  
        	&  Illogical branching condition & 10        \\ 
    logic\_comparison &
        $!=,<,>,<=,>=,==$  & Mistaken logical operators   &    3     \\ 
    logic\_value &
        True, False, None, not None   & Mistaken logical condition &  1   \\ 
    math\_operator &
        +, -, *, /   & Wrong use of mathematical operators & 17       \\ 
\hline
    \multicolumn{3}{|r|}{\textbf{Total}}%
		& 38  \\ 
\hline	
\end{tabular}
\vspace{0.6cm}
\end{center}
\end{table*}

\vspace{-0.1cm}
\subsection{Experimental setup}

In this study, we generated different timeseries of the principal lunar semi-diurnal (M2) tidal constituent with random noises and configurations. The period of M2 astronomical tidal constituent is about 12 hours and 25.2 minutes \cite{hicks2006}. Only one tidal constituent was used so that all MRs are applicable. The length of timeseries varies randomly from 1 week to 1 month at one-hour intervals. The amplitudes and phases were randomly chosen in the range of $[0.1,3]$ metres and $[0^\circ,360^\circ)$, respectively. We generated and used 100 different random datasets (meaning 100 source test cases). 

The input to UTide is the timeseries of tide data and the configuration for the analysis. We configured the program in the ``automation'' mode using the OLS fitting algorithm with a trend for MR1, MR2, MR3, MR4 and MR5 and MR7. For MR6, the program was customized with the given tidal constituent (M2) without a trend in accordance with Equation~\ref{equation:mr6}. Unless prefixing the output with a single tidal constituent, the program returns the output which consists a list of tidal amplitudes and phases that have significant impacts to the signals. We mainly used these outputs to determine whether the MRs are satisfied or violated. For each MR, we ran an experiment twice, one with the source (random) test case and another with the follow-up ones. We repeated the experiments 100 times using 100 different datasets generated above.

\vspace{-0.1cm}
\subsection{Assessment}

To determine the violation or satisfaction of an MR, we examined the relation between the source and follow-up outputs. It involves two conditions.

\begin{enumerate}
  \item For the first condition, we determined whether or not the tidal constituents are consistent with the configuration. All MRs require that the tidal constituents in both the source and follow-up outputs are the same. If they differ by a tidal constituent, the MRs will be considered as violated.
  \item For the second condition, we continued to examine the relation between source and follow-up outputs if the first condition is satisfied. For example, in MR1, we checked whether or not both conditions $\hat A_f=\hat A_s$ and $\hat \phi_f=\hat \phi_s$ are satisfied within a small round-off error (which was set as 0.01 in this study). If one of them is not satisfied, the MR will be considered as violated. 
\end{enumerate}

To measure the effectiveness of our method, we used the mutation score $MS$, which is defined as
\begin{equation}\label{equation:score}
 MS = \frac{M_{K}}{M_{N}}\times 100\%
\end{equation}
where $M_K$ is the total number of mutants that are killed by an MR, $M_N$ is the total number of non-equivalent mutants. 

\vspace{0.3cm}
\section{Result}\label{sec:result}

\subsection{RQ1: Detection of real-life defect}

The proposed MRs were first applied to test the UTide. We observed that MR1 is able to reveal the real-life defect in the software under test. 

With MR1, we found that adding a new point to the timeseries will cause the TAP to produce a totally different output. The detection of error is described as follows. Assume $\hat A_s$ and $\hat\phi_s$ are the amplitude and the phase of a tide component in the timeseries consisting of $M$ points $I_s = \{(t_1,y_1),(t_2,y_2),\ldots,(t_M,y_M)\}$, respectively. The follow-up test case consists of $M+1$ points $I_f = \{(t_1,y_1),$ $(t_2,y_2),\ldots,(t_M,y_M),(t_{M+1},\hat y_{M+1})\}$ where $\hat y_{M+1}$ is
obtai-ned from the prediction mode using Equation~\ref{equation:responsemodel3} with the known $\hat A_s, \hat\phi_s$ and $t_{M+1}$. The follow-up output consists of $\hat A_f$ and $\hat\phi_f$. As required by our MR, the follow-up output should be the same for both amplitude and phase, that is $\hat A_f=\hat A_s$ and $\phi_f=\hat \phi_s$. However, the follow-up phase $\hat\phi_f$ turned out to look like a random number (and hence not equal to $\hat\phi_s$), and related to the choice of $t_{M+1}$, which should not be the case.   

To make sure that there is no human error in the data processing, we have rigorously tracked down every step in the data pipeline. It could be confirmed that the transformation that does not alter the length of timeseries (e.g., scaling, shifting) have not led to any similar abnormal output. The adoption of other MRs (MR2, MR3, MR4, MR5, MR6 and MR7) did not detect any violation. We were yet to determine the root cause of the fault. We anticipated that it could be attributed to how the program converts the computed coefficients (e.g., $\hat B_k$ and $\hat C_k$) to the phase ($\hat \phi_k$), or the mistake in handling the last time step for determining the correct phase. Further investigation will be conducted to locate the buggy statements. 

\subsection{RQ2: Applicability of MT}

Since the defect happens only when we modify the length of timeseries, the UTide is arguably still usable for the mutation analysis. The violation of MR1 is determined from examining only the relation among amplitudes. For other six MRs, both amplitudes and phases were used to derive the mutation scores. Table~\ref{table:results1} presents the summary of experimental results on the mutation analysis.

\begin{table*}[tbp] 
\vspace{0.3cm}
\caption{Mutation scores after 100 test cases.}
\begin{center}
{
\begin{tabular}{|r|rrrrrrr|}
\hline
\textbf{Dataset}
	& \textbf{MR1} 
	& \textbf{MR2}
	& \textbf{MR3} 
	& \textbf{MR4}
	& \textbf{MR5}
	& \textbf{MR6}
	& \textbf{MR7}\\
\hline	
\textbf{Mutation score $MS$} & 24\% & 16\% & 11\% & 16\% & 11\% & 11\% & 13\% \\
\hline
\end{tabular}
}
\label{table:results1}
\end{center}
\end{table*}

It is found that all MRs are able to kill mutants. In other words, all MRs are effective in detecting buggy programs. The mutation scores range from 13\% to 24\% (Table~\ref{table:results1}), subject to the MRs and test cases used. 
The adoption of domain-specific MRs (MR6 and MR7) also allowed us to detect the mutants ($ap=np.hstack((m[:nNR];m[2*nNR : 2+nNR+nR]))$ and $Yu=np.real(ap-am)$) that are unable to be killed by the regression-based MRs (as shown in Table~\ref{table:results2}). Whilst our MRs can detect mutants in the $math\_operator$ and $condition\_if$ categories (Table~\ref{table:results1}), some mutants associated with $logic\_comparison$ and $array\_index$ have yet been killed in these experiments (Table~\ref{table:mutants}).

\begin{table*}[tbp] 
\vspace{0.5cm}
\caption{List of non-equivalent mutants killed and the numbers of times they are killed over the total 100 random test cases.}
\begin{center}
{
\scalebox{0.92}{
\begin{tabular}{|l|l|l|l|rrrrrrr|}
\hline
\textbf{Mutation category} 
	& \textbf{Line} 
	& \textbf{Original} 
	& \textbf{Mutant}
	& \textbf{MR1} 
	& \textbf{MR2}
	& \textbf{MR3} 
	& \textbf{MR4}
	& \textbf{MR5}
	& \textbf{MR6}
	& \textbf{MR7}
	\\
\hline
math\_operator & 242
    & $opt[``rmin"]/(24*lor);$
	& $opt[``rmin"]/(24+lor);$
	& 2 & 0 & 0 & 0 & 0 & 0 & 0 \\
condition\_if & 302
    & $if~not~opt[``notrend"]:$
    & $if~True:$
	& 1 & 0 & 0 & 0 & 0 & 0 & 0 \\	
math\_operator & 338
    & $ap=np.hstack((m[:nNR];$ & $ap=np.hstack((m[:nNR];$
	&  &  &  &  &  &  &  \\
& 
    & $m[2*nNR : 2+nNR*nR]))$ & $m[2*nNR : 2+nNR+nR]))$
	& 0 & 0 & 0 & 0 & 0 & 100 & 100 \\
math\_operator & 342
    & $Xu=np.real(ap+am)$
	& $Xu=np.real(ap-am)$
	& 85 & 0 & 0 & 0 & 0 & 0 & 0 \\
math\_operator & 342
    & $Xu=np.real(ap+am)$
	& $Xu=np.real(ap*am)$
	& 89 & 100 & 0 & 100 & 0 & 81 & 0 \\
math\_operator & 343
    & $Yu=-np.real(ap-am)$
	& $Yu=np.real(ap-am)$
	& 0 & 0 & 0 & 0 & 0 & 100 & 0  \\
math\_operator & 342
    & $Yu=-np.real(ap-am)$
	& $Yu=-np.real(ap+am)$
	& 86 & 14 & 85 & 88 & 93 & 0 & 0 \\
math\_operator & 343
    & $Yu=-np.real(ap-am)$
	& $Yu=-np.real(ap*am)$
	& 86 & 100 & 96 & 82 & 93 & 0 & 22 \\
math\_operator & 343
    & $Yu=-np.real(ap-am)$
	& $Yu=-np.real(ap/am)$
	& 100 & 100 & 0 & 100 & 0 & 0 & 23 \\
condition\_if & 357
    & $if~opt[``twodim"]:$
    & $if~True:$
	& 100 & 100 & 100 & 100 & 100 & 100 & 100 \\
condition\_if & 367
    & $if~opt[``notrend"]:$
    & $if~True:$
	& 100 & 100 & 100 & 100 & 100 & 0 & 100 \\
	\hline	
\end{tabular}
} 
}
\label{table:results2}
\vspace{0.5cm}
\end{center}
\end{table*}

\subsection{RQ3: Effectiveness of MRs}

We also found that the effectiveness of each MR is not the same. 
MR1 has the best performance among all MRs: It could kill about one quarter of the mutants  (Table~\ref{table:results1}).
It is associated with inserting the predicted point into the analysis. The second best MRs include MR2 and MR4, which are related to the reflecting and scaling of sea level, respectively. They had the same mutation score of 16\% (Table~\ref{table:results1}). 

Most detected mutants were about the modification of coefficients through the arithmetic operators ($+,-,*,/$) in preparation for the fitting (Table~\ref{table:results2}). There are three mutants ($opt[``rmin"]/(24+lor);$, $if~True:$ (Line 302), and $Xu=np.real(ap-am)$) that were killed only by MR1. The results suggest that the diversity of MRs is required for detecting different types of faults.

The wide range of effectiveness indicates that some MRs are better than others in detecting faults. As a result, selecting the most effective MR is one key issue for the testing of the TAP software, similar to other types of software.

\section{Discussion}
\label{sec:discuss}

To further improve the effectiveness and robustness of testing, we can consider the following tasks in our future work:
\begin{itemize}
    \item {\it Improvements}: 
    Bugs in the SUT should be detected and corrected before the mutation analysis. We should also need to include more code in the mutation analysis, and at the same time, enhance the diversity of mutants. A better refinement and filtering of non-equivalent mutants are of necessity. 
    \item {\it Extensions}: New MRs are necessary to improve the effectiveness of testing. In addition, the adoption of realistic tide records instead of random data is useful to measure the performance of the method. More TAP programs (other than UTide) are needed to examine the effectiveness of MT.
\end{itemize}

Looking forward, we suggest that MT can be adopted in different ways as follows:
\begin{itemize}
    \item {\it Verification of the TAP software}: Our method is applicable to test the TAP software, either well-established or newly developed ones.
    \item {\it Validation of the TAP data quality and output}: It is very common that a short (1-day to two-week) timeseries of sea level are used. Our MRs can be organized to construct suitable metrics for evaluating the reliability of used data and the confidence of results. 
    \item {\it Verification and validation of high-resolution tide models}: In practice, tidal simulations are mainly validated by means of root mean square errors and correlation coefficients. Since tidal waves simulated in ocean modelling systems are periodic, the domain-specific MRs can be extended to verify and validate them in a similar way.
\end{itemize}

\section{Related works}
\label{sec:related}

In ocean science, Hiremath {\it et al.} \cite{hiremath2021} conducted the first study to adopt MT to validate ocean systems. They developed a machine learning algorithm that is able to identify MRs automatically by optimizing a predefined cost function. They have adopted the method to test a simplified model for computing the kinetic energy of the sea level elevation. 

In the hydraulic field, MT has been successfully applied to test the storm water management model (SWMM) systems \cite{lin2018} and dynamical hydraulic systems model \cite{yang2021wrr}. Lin~{\it et al.} \cite{lin2018} have proposed four MRs related to the change of correlation coefficients ($R^2$), which help detect real-life defects in the SWMM system that is used to simulate the dynamic rainfall-runoff models for estimating the storm water runoffs in urban areas. The work of Yang and Chui \cite{yang2021wrr} focuses on testing machine learning models which predict flood events in Germany. They have proposed three MRs and adopted them to test 13 machine learning models. They then proposed that MT can be further extended to test the general process-based models.

In summary, whilst MT has been shown to be applicable to ocean science and related fields, to the best of our knowledge, no comprehensive work has been carried out to systematically apply MT into the testing of realistic ocean software and modelling systems. 

\section{Conclusion}\label{sec:conclusion}

In this study, we illustrated how to use MT to test the ocean software through a case study of TAP software. In doing so, we have defined a set of MRs, either by adopting MRs derived from its mathematical framework, or by identifying new MRs based on TAP's innate properties. The experiment showed that our MRs were effective in both revealing the real-life fault as well as killing non-equivalent mutants. We also discussed on the future directions to improve and extend the work. The success of our study points out a promising direction for the testing of complex ocean modelling systems, including Ocean General Circulation Models (OGCM) and real-time/operational Ocean Forecast Systems (OFS).

In addition to demonstrating the potentials of MT in testing ocean software, this study also highlights the vision of MT's application into various scientific and industrial simulation systems. In the underwater research, submarine simulators (e.g., UWSim, NAUTIS Maritime Simulator) are used to test the underwater robots including Autonomous Underwater Vehicles (AUVs) for a better development and operation of a submarine. On one hand, submarine simulation might require the inputs from ocean and other simulations; on the other hand, it has features and functions for navigating AUVs. Traffic simulations, as another example, involve complex mathematical formulas and rules for predicting potential traffic flows and events. The work in testing different traffic simulators (such as SUMO and PTV Vissim) and autonomous driving simulators and platforms (e.g., Apollo, Autoware, CARLA, SVL) not only helps improve the safety and the road users' experience, but also may promote the reliability of other vehicular systems including submarines and aeroplanes. It would be worthwhile to investigate how to define MRs (similar to our regression-based MRs). Other application domains of MT in scientific simulations include power distribution systems and even nuclear plants, where a series of MRs can be constructed and used to test the correctness of simulation results such that potential safety-critical errors can be detected and avoided before the real systems are deployed.

\vspace{0.2cm}

\bibliographystyle{ACM-Reference-Format}
\bibliography{references}

\end{document}